\documentclass[12pt,a4paper]{article}
\usepackage{mathrsfs}
\usepackage{epsfig}
\begin{document}
\title{ Bottom partner $B'$ and $Zb$ production at the $LHC$}
\author{ Chong-Xing Yue, Qing-Guo Zeng, Qiu-Yang Shi, Meng-Ying Liao\\
{\small Department of Physics, Liaoning  Normal University, Dalian
116029, P. R. China}
\thanks{E-mail:cxyue@lnnu.edu.cn}}
\date{\today}

\maketitle
\begin{abstract}
Some new physics models, such as "beautiful mirrors" scenario, predict the existence of the bottom partner $B'$. Considering the constraints from the data for the $Z\rightarrow b\overline{b}$ branching ratio $R_{b}$ and the $FB$ asymmetry $A_{FB}^{b}$ on the relevant free parameters, we calculate the contributions of $B'$ to the cross section $\sigma(Zb)$ and the $Z$ polarization asymmetry $A_{Z}$ for $Zb$ production at the $LHC$. We find that the bottom partner $B'$ can generate significant corrections to $\sigma(Zb)$ and $A_{Z}$, which might be detected in near future.

\vspace{0.5cm} \vspace{2.0cm} \noindent
 {\bf PACS numbers}: 12.60.-i, 14.65.Fy, 13.85.Lg

\end{abstract}
\newpage
\noindent{\bf 1. Introduction }\vspace{0.5cm}

Over the past several decades, the standard model $(SM)$ has provided a
consistent description of particles physics and is tested to
per-mille precision by experimented data. Recently, the $ATLAS$ and $CMS$
collaborations have independently reported the discovery [1] of a
neutral scalar particle that seems consistent with the $SM$ Higgs
boson with a mass of about $125\sim 126GeV$. However, some observables related
to the sector of third generation quarks have been observed large
deviations from their $SM$ predictions. The first is the
forward-backward $(FB)$ asymmetry of the bottom-quarks, $A_{FB}^{b}$, which differs by about $2.5\sigma$ deviation from the $SM$ value
at the Z boson pole according the recent global fit result [2]. The second is the $FB$ asymmetry $A_{FB}^{t}$
in top quark pairs produced at the Tevatron, which has larger value than the $SM$
prediction [3]. Furthermore, a recent calculation of  the
$Z\rightarrow b\overline{b}$ branching ratio $R_{b}$,  which includes
new two-loop electroweak corrections, now puts the prediction
in tension with the measured value [4].

It is well known that the top loop in the $SM$ is the largest
contribution to the Higgs mass quadratic divergence. Thus, for the
new physics models to solve the fine tuning problem, there must be
some new particles constrained by symmetry, which cancel this loop. Most
of these new physics models should contain a heavy particle which
shares the gauge quantum numbers of the top quark, generally called
"top partner" [5]. This new particle should be in an electroweak
doublet in order to properly cancel the divergences to the Higgs
mass produced by the top loop. So, this kind of new physics models beyond
the $SM$ predicts the existence of the heavy partner $B'$ of the bottom
quark. Furthermore, if the top and bottom partners have the same mass
hierarchy as the $SM$ top and bottom, the new quark $B'$ may be the first to
be discovered, which has began to be searched at the Tevatron and LHC [6].

Production of the electroweak gauge boson $Z$ associated with a bottom
quark at the $LHC$ is an important background process not only to
Higgs boson production and single top production, but also to
the search for signals of new physics beyond the $SM$, which has been
calculated at next-to-leading order ($NLO$) [7]. Recently, Ref.[8] has defined
the $Z$ polarization asymmetry $A_{Z}$ in the subprocess
$gb\rightarrow Zb$ at the $LHC$ and has shown that $A_{Z}$ is strictly
connected to the $FB$ asymmetry $A_{FB}^{b}$ and is almost free from the
theoretical uncertainties related to $QCD$ scale and parton
distribution function ($PDF$) set variations.

Considering the constraints of the data from $LEP$ for the
$Z\rightarrow b\overline{b}$ branching ratio $R_{b}$ and the $FB$ asymmetry
 $A_{FB}^{b}$ [9] on the $Zb\overline{b}$ couplings $g_{L}^{b}$ and
$g_{R}^{b}$, we are model-independent of calculating the contributions
of the new physics beyond the $SM$ to $Zb$ production at the $LHC$ in
section 2. We find that the correction terms $\delta g_{L}^{b}$ and
$\delta g_{R}^{b}$ generated by new physics cannot give significant
contributions to the production cross section $\sigma(Zb)$. While it is
not this case for the $Z$ polarization asymmetry $A_{Z}$. In section 3, we
study the correction effects of the bottom partner $B'$ on the production cross section $\sigma(Zb)$ and the $Z$ polarization asymmetry
$A_{Z}$. Our numerical results show that, with reasonable values of the relevant free parameters,  $B'$ can generate large
corrections to  $\sigma(Zb)$ and $A_{Z}$. Our conclusion is given in section 4.

\vspace{0.5cm} \noindent{\bf 2. The new physics and $Zb$ production at
the $LHC$ }

\vspace{0.5cm}For the 5-flavor scheme [10], production of the
electroweak gauge boson $Z$ associated with a bottom quark at the $LHC$
proceed via two Feynman diagrams with b-quark exchange in the
s-channel and the t-channel at leading order. Its production cross
section $\sigma(Zb)$ is proportional to the factor
$[(g_{L}^{b})^{2}+(g_{R}^{b})^{2}]$. Thus, new physics can produce
contributions to $\sigma(Zb)$ via correcting the $Zb\overline{b}$
couplings $g_{L}^{b}$ and $g_{R}^{b}$.

The effective $Zb\overline{b}$ couplings can be parameterized by the
Lagrangian
\begin{eqnarray}
{{\cal
L}=\frac{e}{S_{W}C_{W}}\overline{b}\gamma^{\mu}[(g_{L}^{b,SM}+\delta g_{L}^{b})P_{L}+(g_{R}^{b,SM}+\delta g_{R}^{b})P_{R}]bZ_{\mu}},
\end{eqnarray}
with $S_{W}=sin\theta_{W}$ and
$C_{W}=cos\theta_{W}$, in which $\theta_{W}$ is the electroweak mixing angle. $P_{L/R}=(1\mp\gamma_{5})/2$ are the chirality projection
operators.  The $SM$ tree-level couplings $g_{L}^{b,SM}$ and $g_{R}^{b,SM}$ can be written as
:$-\frac{1}{2}+\frac{1}{3}S_{W}^{2}$ and $\frac{1}{3}S_{W}^{2}$, respectively. $\delta g_{L}^{b}$ and $\delta g_{R}^{b}$ represent the new physics
contributions to the $Zb\overline{b}$ couplings. In principle, the  corrections of new physics to the
$Zb\overline{b}$ vertex may give rise to one magnetic moment-type
form factor, proportional to $\sigma^{\mu\nu}q_{\nu}$. However, its
contributions to the $Z\rightarrow b\overline{b}$ branching ratio
$R_{b}$ and the $FB$ asymmetry $A_{FB}^{b}$ are very small and thus have
been neglected in above equation.

\begin{figure}[htb]
\vspace{-0.5cm}
\begin{center}
\epsfig{file=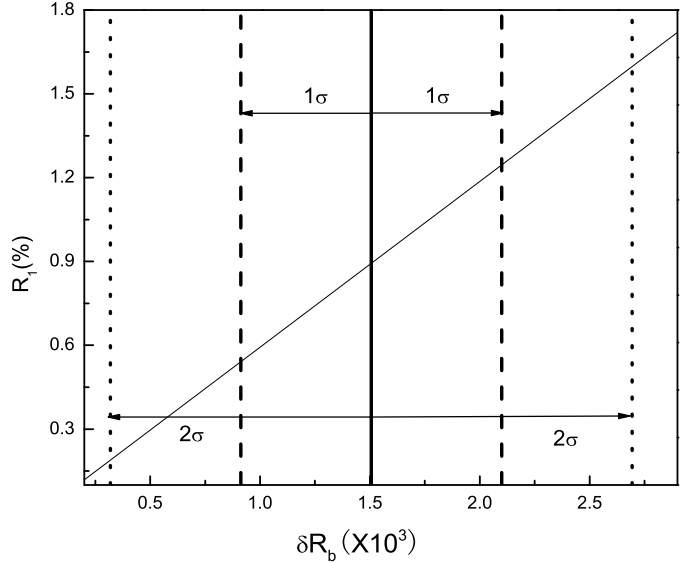,width=300pt,height=260pt}
\vspace{-0.5cm}\caption{The relative correction parameter $R_{1}$ is presented as a function of $\delta R_{b}$.
The \hspace*{1.7cm} regions between dashed lines and between dotted lines correspond $1\sigma$\ and $2\sigma$
\hspace*{1.7cm} allowed regions from $R_{b}$ constraints, respectively. } \label{ee}
\end{center}
\end{figure}

The relative corrections of new physics to $R_{b}^{SM}$ and $A_{FB}^{b,SM}$ can
be approximately written as [11]
\begin{eqnarray}
{\frac{\delta R_{b}}{R_{b}^{SM}}\simeq 2(1-R_{b}^{SM})\frac{g_{L}^{b,SM}\delta g_{L}^{b}+g_{R}^{b,SM}\delta g_{R}^{b}}{(g_{L}^{b,SM})^{2}+(g_{R}^{b,SM})^{2}}},
\end{eqnarray}
\begin{eqnarray}
{\frac{\delta A_{FB}^{b}}{A_{FB}^{b,SM}}\simeq \frac{4(g_{L}^{b,SM})^{2}(g_{R}^{b,SM})^{2}}{(g_{L}^{b,SM})^{4}-(g_{R}^{b,SM})^{4}}(\frac{\delta g_{L}^{b}}{g_{L}^{b,SM}}-\frac{\delta g_{R}^{b}}{g_{R}^{b,SM}})}.
\end{eqnarray}
Where $\delta R_{b}=R_{b}^{exp}-R_{b}^{SM}$ and
$\delta A_{FB}^{b}=A_{FB}^{b,exp}-A_{FB}^{b,SM}$. In above
equations, we have neglected the new physics corrections to the
$Ze\overline{e}$ couplings $g_{L}^{e}$ and $g_{R}^{e}$. The
experimental results  for $R_{b}$ and $A_{FB}^{b}$
are [9]
\begin{eqnarray}
R_{b}^{exp}=0.21629\pm0.00066, ~~ A_{FB}^{b,exp}=0.0992\pm0.0016.
\end{eqnarray}
The recent $SM$ prediction for $R_{b}$, including electroweak two-loop and QCD three-loop
corrections is $R_{b}^{SM}=0.21474\pm0.00003$, which deviates by $2.4\sigma$ deviations below the experimental measured value [2, 4], while the recent global fit result for $A_{FB}^{b}$ is $A_{FB}^{b,SM}=0.1032^{+0.0004}_{-0.0006}$, which is still above the experimental measured value by $2.5\sigma$ deviations [2].

Using above experimental and $SM$ prediction values, one can easily obtain the constraints of the
electroweak precision data on the new $Zb\overline{b}$ couplings
$\delta g_{L}^{b}$ and $\delta g_{R}^{b}$. It is obvious that the data
favor small corrections to $\delta g_{L}^{b}$ and more large shifts
in $\delta g_{R}^{b}$. Considering the discovery of a Higgs-like particle at the $LHC$, Ref. [12] has updated the constraints of the
electroweak precision data on $\delta g_{L}^{b}$ and $\delta g_{R}^{b}$ and there is
\begin{eqnarray}
\delta g_{L}^{b}=0.001\pm0.001, ~~ \delta g_{R}^{b+}=0.016\pm0.005, ~~ \delta g_{R}^{b-}=-0.17\pm0.05.
\end{eqnarray}

We use the relative correction parameter
$R_{1}=[\sigma(Zb)-\sigma^{SM}(Zb)]/\sigma^{SM}(Zb)$ to describe the
corrections of the new $Zb\overline{b}$ couplings
$\delta g_{L}^{b}$ and $\delta g_{R}^{b}$ to the cross section of the process $pp\rightarrow Zb$, in which $\sigma(Zb)$ denotes the total production cross section including the contributions from the $SM$, $\delta g_{L}^{b}$, and $\delta g_{R}^{b}$.
In our calculations, the $PDFs$ of the bottom quark and gluon are taken as the $CTEQ6L$ $PDFs$ [13] with renormalization and factorization scales $\mu_{R}=\mu_{F}=M_{Z}$. To make our numerical results more realistic, we have applied the cuts on the $b-jet$ with transverse momentum $P_{T}>15GeV$
 and a rapidity range $|\eta|<2$.
It is obvious that the radiative corrections to $\sigma(Zb)$ and $\sigma^{SM}(Zb)$ are canceled in the relative correction parameter $R_{1}$.
In Fig.1 we plot $R_{1}$ as a function of $\delta R_{b}$ for $1\sigma$ and $2\sigma$ constraints from the $R_{b}$ experimental value. One can see that the value of $R_{1}$ allowed by the $R_{b}$ constraints is very small.
For the theory value of $R_{b}$ being consistent with its experimental value with $1\sigma$ and $2\sigma$ error bars, the values of the parameter $R_{1}$ are in the ranges of $0.53\%\sim1.3\% $ and $0.14\%\sim1.7\%$, respectively, which are much smaller than the $QCD$ corrections [7].

Searching for the gauge boson $Z$ produced in association with the bottom quark has been performed at the $LHC$. Recently, the $ATLAS$ collaboration [14] has reported their measurement of the $Zb$ production cross section and found that it is in good agreement with the $SM$ prediction including the $NLO$ $QCD$ corrections. Considering the statistical and systematic uncertainties, the $ATLAS$ data cannot give severe constraints on the new $Zb\overline{b}$ couplings $\delta g_{L}^{b}$ and $\delta g_{R}^{b}$.

\begin{figure}[htb]
\vspace{-0.5cm}
\begin{center}
 \epsfig{file=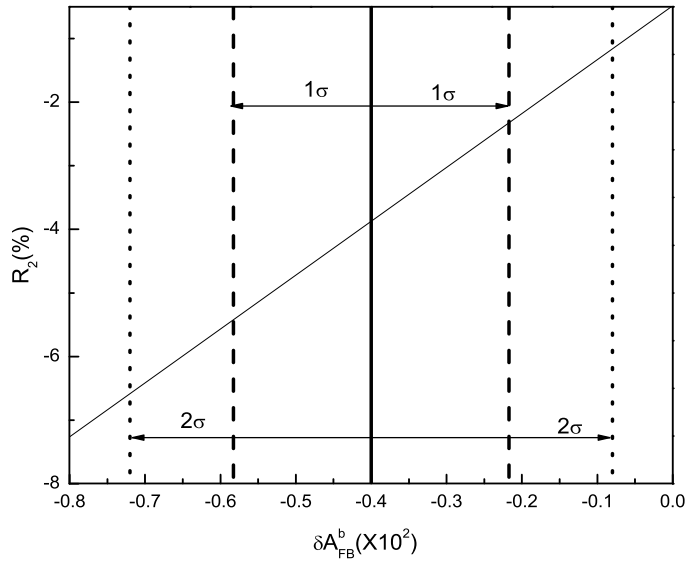,width=300pt,height=260pt}
\vspace{-0.5cm} \caption{  The relative correction parameter $R_{2}$ as a function of $\delta A_{FB}^{b}$. The regions \hspace*{1.7cm} between  dashed lines and between dotted lines correspond $1\sigma$ and $2\sigma$ allowed \hspace*{1.7cm} regions from  $A_{FB}^{b}$ constraints, respectively. } \label{ee}
\end{center}
\end{figure}

Compared to the cross section, decay width, etc, the asymmetry, which is defined as a ratio of observables, is not sensitive to the theoretical uncertainties. The asymmetry can be utilized to study the detail properties of the particles and further to investigate underlying dynamics in and/or beyond the $SM$. Measurement of the asymmetry at the $LEP$ and Tevatron has provided rich informations about the $SM$ and various new physics models.

The $Z$ polarization asymmetry $A_{Z}$ in $Zb$ production at the $LHC$ can be defined as
\begin{eqnarray}
{A_{Z}=\frac{\sigma(Z_{R}b)-\sigma(Z_{L}b)}{\sigma(Z_{R}b)+\sigma(Z_{L}b)}},
\end{eqnarray}
where $\sigma(Z_{R}b)$ and $\sigma(Z_{L}b)$ are the hadronic cross sections of $Z_{R}b$ and $ Z_{L}b$ production at the LHC, respectively. Ref.[8] has shown that $A_{Z}$ is connected to the $Zb\overline{b}$ $FB$ asymmetry $A_{FB}^{b}$ and given its $SM$ prediction value. If the large deviation between the $SM$ prediction and the $LEP$ measurement of $A_{FB}^{b}$ indeed exists and comes from the new $Zb\overline{b}$ couplings $\delta g_{L}^{b}$ and $\delta g_{R}^{b}$, then these new couplings should generate significant contributions to $A_{Z}$.

To see whether the correction effects of the new $Zb\overline{b}$ couplings $\delta g_{L}^{b}$ and $\delta g_{R}^{b}$ on the $Z$ polarization asymmetry $A_{Z}$ can be detected at the $LHC$, we define the relative correction parameter $R_{2}=\delta A_{Z}/A_{Z}^{SM}$ with $\delta A_{Z}= A_{Z}^{total}-A_{Z}^{SM}$. Our numerical results are shown in Fig.2, in which we plot $R_{2}$ as a function $\delta A_{FB}^{b}$ to consistent with the experimental value of $A_{FB}^{b}$ with $1\sigma$ and $2\sigma$ error bars. One can see that the absolute value of $R_{2}$ can reach $6.8\%$. Considering $A_{Z}$ almost free from the theoretical uncertainties, we hope that the $LHC$ might detect this correction effects and confirm or obviate the $A_{FB}^{b}$ anomaly.

\vspace{0.5cm} \noindent{\bf 3. The bottom partner $B'$ and $Zb$ production at the $LHC$ }

\vspace{0.5cm} So far, the $Zb\overline{b}$ $FB$ asymmetry $A_{FB}^{b}$ measured in $Z$ boson decays at $LEP$ experiments still exist $2.5\sigma$ deviations from the $SM$ prediction [2]. Considering modification of the $SM$ $Zb\overline{b}$ couplings $g_{L}^{b,SM}$ and $g_{R}^{b,SM}$, some new physics models have been proposed to cure the large discrepancy [15, 16, 17]. Ref. [17] proposed the beautiful mirrors model, which introduces vector-like quarks which mix with the bottom quark subtly affecting its couplings to the gauge boson $Z$ and addressing the observed anomaly in $A_{FB}^{b}$. This model predicts the existence of the bottom partner  $B'$. Some of their phenomenological consequences have been explored in Refs. [17, 18]. Taking into account of the constraints on the relevant free parameters from explaining the current $R_{b}$ and $A_{FB}^{b}$ deviations [2, 4, 12], we consider the contributions of the bottom partner $B'$ to the hadronic cross section $\sigma(Zb)$ and the $Z$ polarization asymmetry $A_{Z}$ for $Zb$ production at the $LHC$ in this section.

The beautiful mirrors model [17] extends the $SM$ by introducing two sets of vector-like quarks, $\psi_{L,R}$ with quantum numbers (3, 2, -5/6) and $\xi_{L,R}$ with quantum numbers (3, 1, -1/3), in which the $SM$ Higgs is the only source of electroweak symmetry breaking ($EWSB$). In terms of its $SU(2)$ components, $\psi_{L,R}$ decomposes as
\begin{eqnarray}
\psi_{L,R}=\left( \begin{array}{c} \omega_{L,R} \\\chi_{L,R}
\end{array}\right),
\end{eqnarray}
where $\omega$ is a charge -1/3 quark and $\chi$ has charge -4/3. It is assumed that the new quarks only couple to the third generation $SM$ quarks, which are governed by the $SU(3)\times SU(2)\times U(1)$ gauge invariance. These new quarks mix with the $SM$ bottom quark to explain the measured value of $A_{FB}^{b}$ and have small mixing with the two lighter $SM$ generation quarks to satisfying the constraints from rare decay processes of the bottom and strange mesons such as $B\rightarrow X_{s}\gamma$, $B\rightarrow l^{+}l^{-}X$, $B\rightarrow J/\Psi K_{s}$ and $K\rightarrow \pi\nu\overline{\nu}$.

In the beautiful mirrors model, the couplings between the gauge boson $Z$ and the down-type quarks may be written in matrix form [17]
\begin{eqnarray}
{\cal
L}_{Z}=\frac{e}{S_{W}C_{W}}\overline{d}\gamma^{\mu}(LP_{L}+RP_{R})d Z_{\mu}+h.c.,
\end{eqnarray}
where $d=(b_{1},b_{2},b_{3})$, in which $b_{1}$ is mainly the $SM$ bottom quark field, $b_{2}$ is mostly $\omega$ and $b_{3}$ is mostly $\xi$. We call $b_{2}$ as bottom partner $B'$ and consider its contributions to $Zb$ production at the $LHC$. The coupling matrices $L$ and $R$ are written as
\begin{eqnarray}
 L=U_{d}^{\dag}g_{L}U_{d}, ~~
 R=W_{d}^{\dag}g_{R}W_{d},
 \end{eqnarray}
where $g_{L}=Diag(-\frac{1}{2}+\frac{1}{3}S_{W}^{2},~~ \frac{1}{2}+\frac{1}{3}S_{W}^{2}, ~ \frac{1}{3}S_{W}^{2})$,~
$g_{R}=Diag(\frac{1}{3}S_{W}^{2}, ~\frac{1}{2}+\frac{1}{3}S_{W}^{2},~ \frac{1}{3}S_{W}^{2})$. The unitary matrices $U_{d}$ and $W_{d}$ transform the left- and right-handed gauge eigenstates into the corresponding mass eigenstates, which can diagonalize the mass matrix,
\begin{eqnarray}
U_{d}^{\dag}M_{d}W_{d}=
\left( \begin{array}{ccc}m_{1}&0
 &0\\0&m_{2}&0\\0&0&m_{3}\end{array}\right),
\end{eqnarray}
where $m_{1}=m_{b}$, $m_{2}$ and $m_{3}$ are the $SM$ bottom quark mass, and two new quark masses. The matrix $U_{d}$ can be parameterized as
\begin{eqnarray}
U_{d}=\left( \begin{array}{ccc}C_{12}^{L}C_{13}^{L}&S_{12}^{L}C_{13}^{L}&S_{13}^{L}\\
-S_{12}^{L}C_{23}^{L}-C_{12}^{L}S_{23}^{L}S_{13}^{L}&C_{12}^{L}C_{23}^{L}-S_{12}^{L}S_{23}^{L}S_{13}^{L}&S_{23}^{L}C_{13}^{L}\\
S_{12}^{L}S_{23}^{L}-C_{12}^{L}C_{23}^{L}S_{13}^{L}&-C_{12}^{L}S_{23}^{L}-S_{12}^{L}C_{23}^{L}S_{13}^{L}&C_{23}^{L}C_{13}^{L}
 \end{array}\right),
\end{eqnarray}
with $C_{12}^{L}=cos\theta_{12}^{L}$ and so on, in which $\theta_{ij}$ are the mixing angles. The matrix $W_{d}$ has an analogous expression but with $\theta_{ij}^{L}\rightarrow \theta_{ij}^{R}$.

Using above equations, one can write the explicit expression forms for the $Zb\overline{b}$, $ZB'\overline{B'}$, $Zb\overline{B'}$ couplings, etc, and further give the correction terms $\delta g_{L}^{b}$ and $\delta g_{R}^{b}$ to the $SM$ $Zb_{L}\overline{b_{L}}$ and $Zb_{R}\overline{b}_{R}$ couplings. To predigest our calculation, we set $S_{12}^{R}=S_{R}\neq0$, $S_{13}^{L}=S_{L}\neq0$, and all other mixing angles equal to zero. In this simply case, the couplings, which are related our calculation, can be written as
\begin{eqnarray}
\delta g_{L}^{b}=\frac{S_{L}^{2}}{2}, ~~  \delta g_{R}^{b}=\frac{S_{R}^{2}}{2};
\end{eqnarray}
\begin{eqnarray}
g_{L}^{bB'}=0,  ~~ g_{R}^{bB'}=-\frac{e}{2S_{W}C_{W}}S_{R}C_{R}.
\end{eqnarray}
Comparing the experimental measured values of the $Z\rightarrow b\overline{b}$ branching ratio
$R_{b}$ and $FB$ asymmetry $A_{FB}^{b}$ with their current theoretical prediction values [2, 4], one can obtain the constraints on the  mixing parameters $S_{L}$ and $S_{R}$. To make $A_{FB}^{b}$ and $R_{b}$ consistent with their experimental measured values with
$1\sigma$ and $2\sigma$ error bars, the mixing parameters $S_{L}$
and $S_{R}$ must satisfy the relation
\begin{eqnarray}
1\sigma: ~~ 0\leq S_{L}^{2}\leq 0.004, ~~ 0.022\leq S_{R}^{2}\leq 0.042;\\
2\sigma: ~~ 0\leq S_{L}^{2}\leq 0.006,~~  0.012\leq S_{R}^{2}\leq0.052.
\end{eqnarray}

\begin{figure}[htb]
\vspace{-0.2cm}
\begin{center}
 \epsfig{file=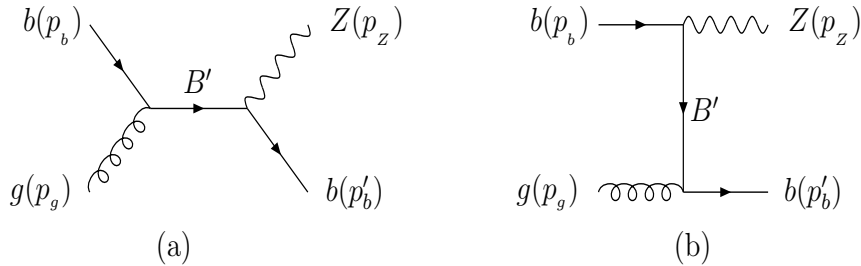,width=320pt,height=100pt}
\vspace{-0.5cm} \caption{ Feynman diagrams for the $B'$ contributions to $Zb$ production at the $LHC$. } \label{ee}
\end{center}
\end{figure}

 The couplings of the $SM$ quarks and new down-type quarks to the Higgs boson $H$ and the gauge boson $W$ can be obtained from Ref. [17].

The couplings of the new fermions to the $SM$ gauge bosons and ordinary fermions are uniquely fixed by gauge invariance [19]. The general Lagrangian describing the interactions between the $SM$ bottom quark, its partner $B'$ and gluon is fixed by $SU(3)$ gauge invariance to be of magnetic moment type [20, 21]
\begin{eqnarray}
{{\cal
L}_{gbB'}=\frac{g_{s}}{2\Lambda}G_{\mu\nu}^{a}\overline{b}\lambda^{a}(K_{L}^{b}P_{L}+K_{R}^{b}P_{R})\sigma^{\mu\nu}B'+h.c.},
\end{eqnarray}
where $G_{\mu\nu}^{a}$ is  the gluon field strength tensor with the
 color index $a=1,...,8$, and $g_{s}$ is the $QCD$ coupling constant,
  $\lambda^{a}$ are the fundamental $SU(3)$ representation matrices.
  In this Letter, we set the new physics scale $\Lambda$ to $M_{B'}$
  and assume that the coupling constants $K_{L}^{b}$ and $K_{R}^{b}$
   are both of order one in the strongly interacting theory. It is should be noted that,
   using this type couplings, Ref.[22] has considered the contributions
    of $B'$ to $tW$ association production and discussed the possibility of detecting the bottom partner $B'$ at the $LHC$.

\begin{figure}[htb]
\vspace{-0.5cm} {\includegraphics[scale=0.68]{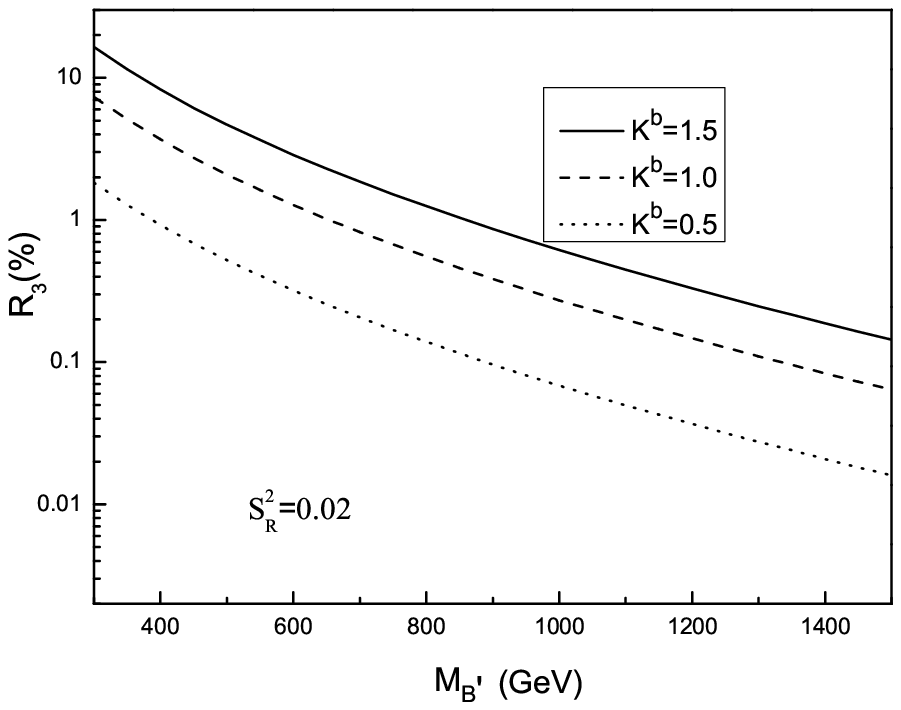}}
{\includegraphics[scale=0.68]{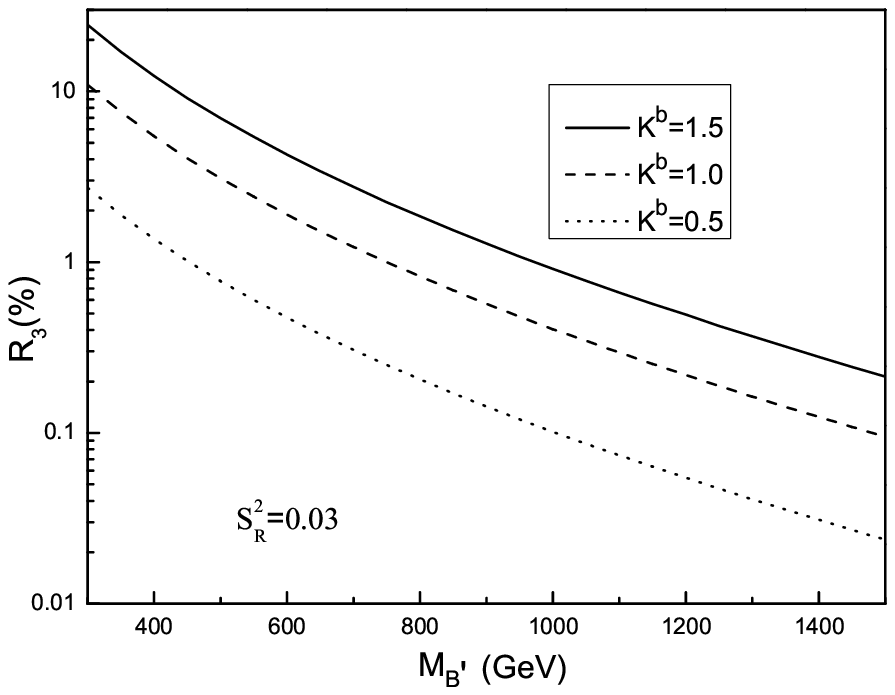}}

{\includegraphics[scale=0.68]{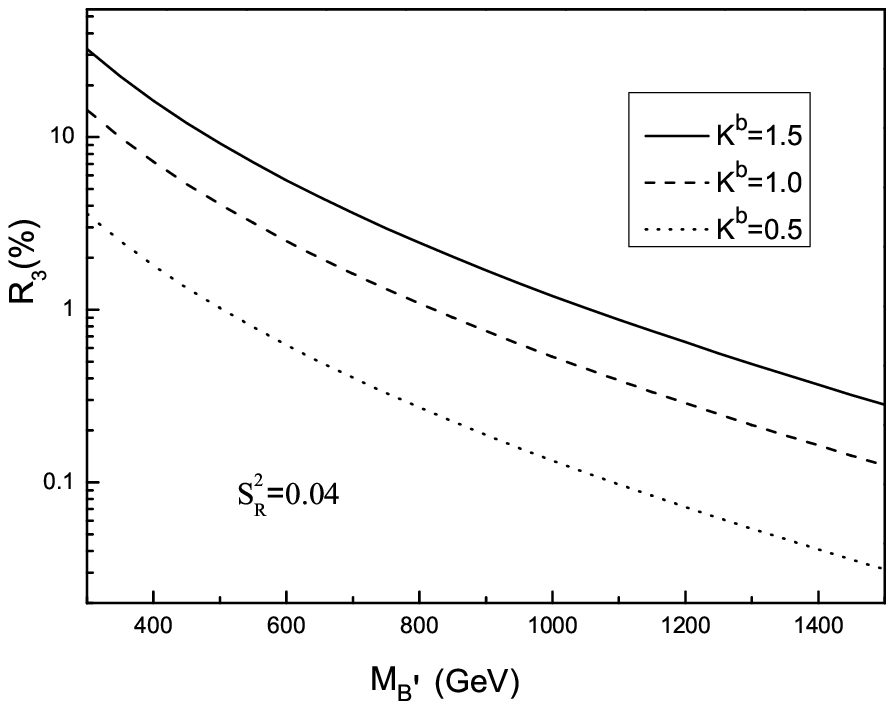}}
{\includegraphics[scale=0.68]{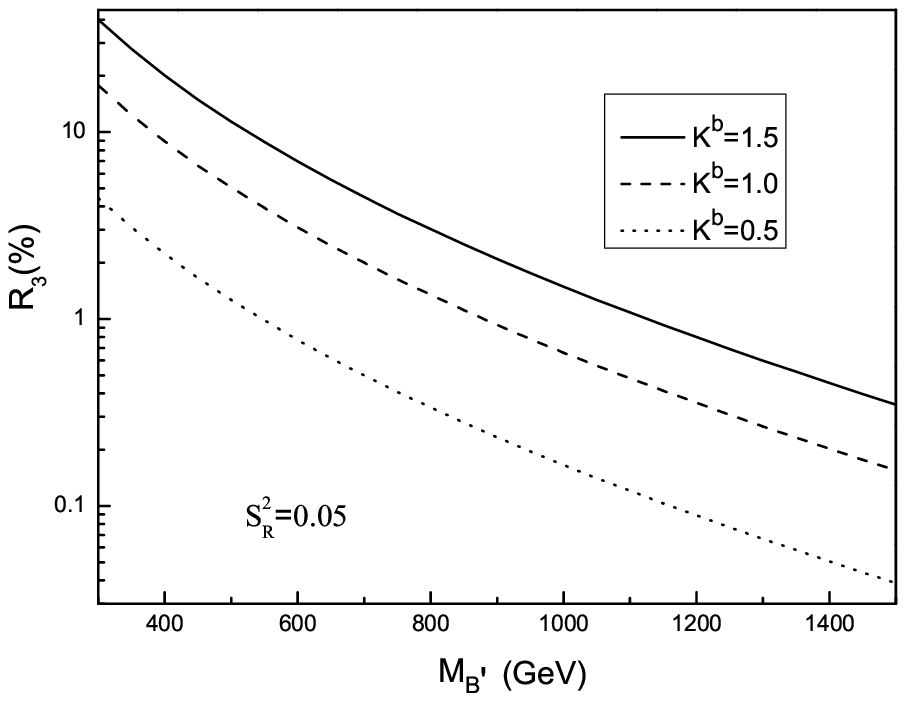}}

\vspace{-0.5cm} \caption{ The relative
 correction parameter $R_{3}$
 as a function of the bottom partner $B'$
\hspace*{1.7cm} mass $M_{B'}$ for different values
 of the free parameters $S_{R}$ and $K^{b}$. }
 \label{ee}
\label{ee}
\end{figure}

From above discussions we can see that the bottom partner $B'$ can
contribute to $Zb$ production at the $LHC$ via s-channel and t-channel
 $B'$ exchanges, as shown in Fig.3. Our numerical results are given in Fig.4, in which we plot the relative
 correction parameter $R_{3}=(\sigma^{total}-\sigma^{SM})/\sigma^{SM}$
 as a function of the bottom partner $B'$ mass $M_{B'}$,  $ \sigma_{total}$  includes
 the contributions from the $SM$ and the bottom partner $B'$.
 Since the contributions of the new couplings $\delta g_{L}^{b}$ and
  $\delta g_{R}^{b}$ to $Zb$ production are very small,
  we have not included their correction effects in the relative correction
  parameter $R_{3}$. In our numerical calculation, we have considered
  the constraints of the electroweak precision measurement,
  such as $R_{b}$ and $A_{FB}^{b}$, on the mixing parameters
  $S_{L}$ and $S_{R}$, and assumed the total decay width
   $\Gamma_{total}(B')=\Gamma(B'\rightarrow tW)+\Gamma(B'\rightarrow Z b)+\Gamma(B'\rightarrow Hb)+\Gamma(B'\rightarrow gb)$
   and $K_{L}^{b}=K_{R}^{b}=K^{b}$.
 One can see from Fig.4 that,  with reasonable values of the relevant free parameters,
 the bottom partner $B'$ can generate significant contributions to $Zb$ production
 at the $LHC$. For the mixing parameter $S_{R}$ consistent with the experimental
 values of $A_{FB}^{b}$ with $1\sigma$ and $2\sigma$ error bars,
  $0.5\leq K^{b}\leq 1.5$ and $300GeV\leq M_{B'}\leq 1500GeV$,
  the values of $R_{3}$ are in the ranges of $1.8\times10^{-4} \sim 0.34$ and $9.7\times10^{-5} \sim 0.41$, respectively. The correction of the bottom partner $B'$ to $Zb$ production at the $LHC$ is comparable to its $NLO$ $QCD$ correction and might be larger than the $NLO$ $QCD$ correction for taking special values of the free parameters.

\begin{figure}[htb]
\vspace{-0.4cm}
\begin{center}
 \epsfig{file=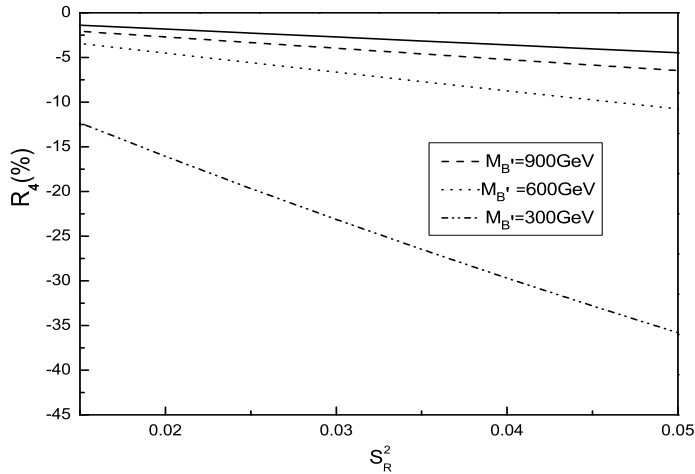,width=300pt,height=220pt}
\vspace{-0.5cm} \caption{The relative  correction parameter $R_{4}$
 as a function of  $S_{R}^{2}$ for $S_{L}^{2}=0.004$,  $K^{b}=1$ \hspace*{1.7cm} and three values of the $B'$ mass $M_{B'}$.
The solid line expresses the contribu-\hspace*{1.7cm} tions  of the new $Zb\overline{b}$ couplings $\delta
g_{L}^{b}$ and $\delta g_{R}^{b}$ and other lines denote the total \hspace*{1.7cm} contributions of  the beautiful mirrors model. } \label{ee}
\end{center}
\end{figure}

In the beautiful mirrors  model, the correction effects on the $Z$
polarization asymmetry $A_{Z}$ for $Zb$ production at the $LHC$ come
from two sources: the new $Zb\overline{b}$ couplings $\delta
g_{L}^{b}$ and $\delta g_{R}^{b}$, and the bottom partner $B'$.  The contributions
  of $B'$ to $A_{Z}$ is not related the free parameter $S_{L}$ and the contributions of
  $\delta g_{L}^{b}$ are much smaller than those for $\delta g_{R}^{b}$ and $B'$,
   so we fix the value of the free parameter $S_{L}$ to $S_{L}^{2}=0.004$.
    The relative corrections of the beautiful mirrors model to $A_{Z}$ is presented by the parameter $R_{4}$,
    which is plotted as a function of $S_{R}^{2}$ for $K^{b}=1$ and three values of the $B'$ mass $M_{B'}$
    in Fig.5. The absolute value of the parameter $R_{4}$ increases as $M_{B'}$ decreases and $S_{R}$ increases.
    For $300GeV \leq M_{B'}\leq 900GeV$ and $0.015 \leq S_{R}^{2}\leq 0.05$, its value is in the range of $-35.8\% \sim -1.4\%$.
    Thus, the possible signatures of the beautiful mirrors model might be detected at the $LHC$ via measuring its correction effects on the $Z$ polarization asymmetry $A_{Z}$ in near future.

\vspace{0.5cm} \noindent{\bf 4. Conclusions }

\vspace{0.5cm}The electroweak precision measurements can generate
severe constraints on the new physics beyond the $SM$.
The large deviation between the $SM$ prediction and
the $LEP$ measurement of the $FB$ asymmetry $A_{FB}^{b}$ and
 the $Z\rightarrow b\overline{b}$ branching
 ratio $R_{b}$ require that the new physics has large corrections to the $SM$ $Zb_{R}\overline{b_{R}}$
  coupling $g_{R}^{b,SM}$ and small corrections to the
   $SM$ $Zb_{L}\overline{ b_{L}}$ coupling $g_{L}^{b,SM}$.
   In this Letter, we first consider the contributions of
   the new $Zb\overline{b}$ couplings $\delta g_{L}^{b}$
   and $\delta g_{R}^{b}$ to the hadronic cross section $\sigma(Zb)$
   and the $Z$ polarization asymmetry $A_{Z}$ for $Zb$ production at
   the $LHC$. We find that the relative correction of $\delta g_{L}^{b}$
   and $\delta g_{R}^{b}$ to $\sigma(Zb)$ is very small, while can reach $6.8\%$ for $A_{Z}$.

Some new physics models beyond the $SM$  predict the existence of the bottom partner $B'$.
  Considering the constraints from the electroweak precision
   measurements on this new physics model, we further calculate
   the contributions of $B'$ to the production cross section $\sigma(Zb)$
   and the $Z$ polarization asymmetry $A_{Z}$. Our numerical results
   show that the "beautiful mirrors" scenario can give significant corrections
   to the physical observables $\sigma(Zb)$ and $A_{Z}$, which might be detected at the $LHC$ in near future.

\section*{Acknowledgments} \hspace{5mm}This work was
supported in part by the National Natural Science Foundation of
China under Grants Nos.10975067 and 11275088, the Natural Science Foundation of the Liaoning Scientific Committee
(No. 201102114), and Foundation of Liaoning Educational Committee (No. LT2011015).

\end{document}